\documentclass[aps,prb,twocolumn,showpacs,preprintnumbers,amsmath,amssymb]{revtex4}

\usepackage{graphicx}
\usepackage{dcolumn}
\usepackage{bm}
\begin{document}


\title{Construction of a Cantilever-Andreev-Tunneling rig and its applications to superconductors}

\author{W. K. Park}
\author{L. H. Greene}
\affiliation{Department of Physics and the Frederick Seitz Materials Research Laboratory, University of Illinois at Urbana-Champaign, Urbana, Illinois 61801, USA}

\date{\today}

\begin{abstract}
A technique for point-contact spectroscopy, based on an electro-mechanical mechanism for the contact formation, has been developed. It is designed to be used in both $^4$He and $^3$He cryostats.
The performance has been demonstrated by conductance measurements on various kinds of superconductors, including the conventional superconductor Nb, the two-band superconductor MgB$_2$, and the heavy-fermion superconductor CeCoIn$_5$. Characteristic conductance spectra obtained prove this technique is useful for the investigation of the superconducting order parameter. Advantages of this technique such as its simplicity and versatility are described.
\end{abstract}


\maketitle

\section{\label{sec:intro}Introduction}

Investigation of the order parameter symmetry and the pairing mechanism is of fundamental importance to understand the physics of superconductors. Tunneling spectroscopy, which utilizes the quantum mechanical tunneling of quasi-particles (QPs) across a tunneling barrier between a normal metal (N) and a superconductor (S), has been widely and successfully adopted to study both conventional\cite{wolf85} and unconventional superconductors.\cite{zasadzinski03} Andreev reflection (AR) spectroscopy has also been playing an important role. This technique relies on the Andreev reflection process where the retro-reflection of a QP occurs uniquely at an N/S interface.\cite{andreev64} Blonder, Tinkham, and Klapwijk (BTK)\cite{btk82} formulated a successful theory to describe the transitional behavior from tunneling to AR as the strength of the tunnel barrier becomes smaller. There have been many extended or modified versions of the BTK theory. For instance, Tanaka and Kashiwaya\cite{kashiwayatanaka} extended the BTK theory to $d$-wave superconductors.

Point-contact spectroscopy (PCS) is a technique used to investigate the electronic transport in nano-scale junctions. The BTK theories have been providing solid foundations for the analysis of PCS data obtained on superconductors.\cite{blonder83} In particular, since the recent discovery of superconductivity in MgB$_2$,\cite{nagamatsu01} PCS has been playing a central role in defining and extracting information on the two energy gaps.\cite{szabo01,gonnelli02} With the simplicity of implementation, PCS has proved to be a useful tool not only to probe the superconducting order parameter but also to investigate bosonic spectrum for the pairing.\cite{jansen80,duif89,naidyuk05}

In order to provide spectroscopic information, the point contact should be in a ballistic (or Sharvin\cite{sharvin65}) limit, i.e., smaller than the electronic mean free paths of either of the contact materials. Depending on the kind of superconductors of interest and the available experimental
setup, various techniques to make ballistic contacts have been developed,\cite{jansen80,duif89,naidyuk05}
including thin films,\cite{yanson74} the spear-anvil technique,\cite{jansen77} pressed In contacts or Ag paint drops,\cite{gonnelli02} and scanning tunneling microscopy (STM).\cite{hawley86}
Each technique has unique advantages and disadvantages. The most common technique is the use of a differential screw,\cite{blonder83} that uses fine mechanical control. This has been adopted extensively since it is straightforward and easy to implement in a cryostat. However, it has the disadvantages of mechanical backlash and difficulty in application to $^3$He or dilution refrigeration systems.

Because certain superconductors such as heavy-fermion superconductors (HFS) have low T$_c$s, requiring low temperature measurements, it is necessary to develop a PCS technique applicable in $^3$He or dilution refrigeration systems.\cite{goll93,heil93} Because of our interest in a new family of HFS, CeMIn$_5$ (M=Co, Rh, Ir), we have developed a new technique for PCS to be used with a $^3$He cryostat. The probe, coined as a Cantilever-Andreev-Tunneling (CAT) rig, is also designed to be used in a liquid $^4$He system.
In the following, we describe the design and construction of the CAT rig and present applications to three different kinds of superconductors: Nb, MgB$_2$, and CeCoIn$_5$.

\section{\label{sec:design}Design and construction of the CAT rig}

The CAT rig is designed to be attached to either a $^3$He cryostat or a $^4$He probe. It is also designed to be used in two modes: Andreev reflection and planar tunneling. In the planar tunneling mode, the tunneling conductance is measured by bringing a normal-metal electrode, flattened and coated with a thin robust insulating layer, into contact with the superconductor. This technique is useful for tunneling into superconductors on which planar tunnel junctions are not easily fabricated. However, since it is not trivial to establish a technique for preparing such a counter-electrode, this mode has not yet been realized. The Andreev reflection or point contact mode is easier to implement, since one only has to fabricate a fine tip and to bring it onto the sample surface. For those reasons mentioned in the previous section, we decided to build a rig based on a two-step electro-mechanical approach, that is, a combination of a coarse approach using a screw mechanism and a fine adjustment using piezoelectric bimorphs.

\begin{figure}
\includegraphics{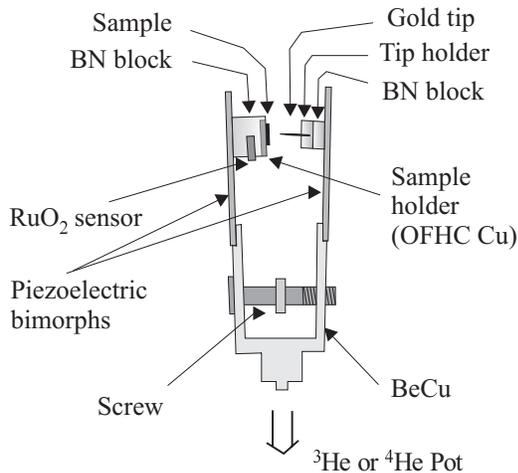}
\caption{\label{fig:CATrig} Schematic of the CAT rig.
The main body is made of BeCu to ensure a reversible movement by the screw. The contact can be adjusted further by the piezoelectric bimorphs. (Not to scale)}
\end{figure}

A schematic of our CAT rig is shown in Fig.~\ref{fig:CATrig}. The body is made of BeCu to ensure enough elasticity for reversible movement of the tip and the sample. It has two arms, with a piezoelectric bimorph attached to the end of each arm. This part is finely machined to be slant to ensure that the sample surface can be aligned parallel to the surface of the counter-electrode when the contact is made in the planar tunneling mode. The piezoelectric bimorphs with nonmagnetic electrodes are driven by a linear power amplifier with a maximum output of $\pm$ 200 V. The input for the power amplifier is provided either by the DAC output of a lock-in amplifier or by the 20-bit DAC board. The former is used for the point contact mode since the 12-bit resolution is sufficient. The latter input is required for the planar tunneling mode to obtain more precise control of the tip-sample distance. In this case, combining the maximum deflection (325 $\mu$m at room temperature and 1/7-1/8 of this value at 4.2 K \cite{chen93}) and the resolution of the DAC (19.1 $\mu$V), the ultimate resolution for the control of the tip-sample distance is estimated to be 1.4 \AA \ at room temperature and 0.2 \AA \ at 4.2 K. In principle, it is therefore possible to achieve control fine enough for the planar tunneling mode.
A boron nitride (BN) rod is machined into rectangular and cylindrical blocks to provide electrical isolation for the sample and the tip holders from the piezoelectric bimorphs. These BN blocks are glued onto the piezoelectric bimorphs using a low temperature epoxy. The tip holder is made of aluminum and machined to screw into the cylindrical BN block. A tip is fastened to the holder by a set screw. Two copper wires are soldered to the tip holder to provide electrical leads. An OFHC Cu plate is the sample holder and is mounted on the BN block with epoxy. Four electrodes are also embedded in this BN block with epoxy. When used in the $^3$He cryostat, the sample holder is thermally anchored by multiple pairs of twisted Cu wires to the $^3$He pot to ensure good thermal conduction. The temperature of the sample is monitored with a RuO$_2$ temperature sensor embedded into the BN block using N-grease.

\begin{figure}[t]
\includegraphics{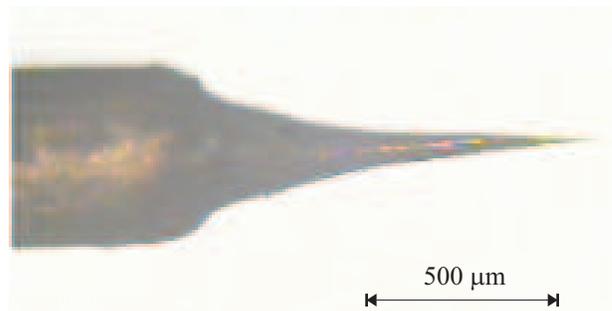}
\caption{\label{fig:AuTip} Optical microscope image of an electrochemically prepared Au tip.}
\end{figure}

Procedures for fabricating fine gold tips to be used in STM and PCS have been well established.\cite{chen93,libioulle95} Figure~\ref{fig:AuTip} shows an optical microscope image of a gold tip prepared using an electrochemical etching technique.\cite{libioulle95} The etching of gold occurs via reduction-oxidation reactions aided by $dc$ bias in an acid solution.\cite{ren04} For the best performance in PCS, the procedure is optimized to produce a thin, long, smooth, uniform and clean gold tip, as shown in Fig.~\ref{fig:AuTip}. A long, thin tip is required because good conductance data are obtained when the tip is bent to make a contact.\cite{blonder83} The tip is fabricated from a high-purity (99.9985\%) gold wire of 500 $\mu$m diameter and 7-8 mm length, which acts as the anode. A thin platinum plate acts as the cathode in the HCl solution. A 5 kHz $dc$ pulse of 10 V and 10 $\mu$s duration is applied between them. The HCl solution of normality 12.1 is heated to 50-70 $^{\circ}$C to facilitate the reaction. Right after etching is finished, the tip is rinsed thoroughly with hot deionized water to remove salts such as AuCl$_4^-$. The smoothness of surface is important to avoid forming multiple contacts and contamination by carbon.\cite{ren04}

\section{\label{sec:exp}Experimental Details}

In this section, we describe detailed procedures for making measurements using the CAT rig in a point contact mode. Since it is designed to fit to both $^3$He and $^4$He cryostats and the procedures differ,
we consider each case separately. In general, the $^4$He system is much easier and simpler to use.
After describing the mounting of the sample and cool-down procedures in each cryostat, we describe how we monitor the sample-tip contact resistance during the cool-down procedure.

\begin{figure}[t]
\includegraphics{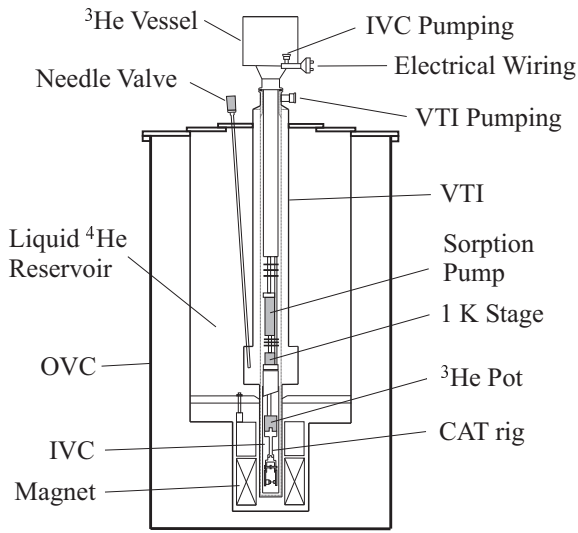}
\caption{\label{fig:CATsetup} Schematic of $^3$He and $^4$He cryostats used for
PCS measurements. (Not to scale)}
\end{figure}

\subsection{$^4$He system}

The detailed procedure for making contacts to the sample when used with $^4$He cryostat is as follows.
The CAT rig is attached to a probe before the sample and the tip are mounted. The sample is then mounted on the sample holder using N-grease (see Fig.~\ref{fig:CATrig}). For the measurement of conductance, two electrical leads are attached to the superconductor using Ag paint. The gold tip is used as the counter-electrode. After the sample and the tip are mounted, a contact is formed by adjusting the screw while being observed via an optical microscope. Typically, good conductance data are obtained when the tip is bent to make a contact.\cite{blonder83} This is presumably because it makes the contact more stable. Once a contact is formed, the protective cover made of OFHC copper sheet is loaded around the CAT rig.

The probe is carefully lifted and then slowly inserted into the variable temperature insert (VTI)
(see Fig.~\ref{fig:CATsetup}), which has been kept above $\sim$ 100 K. An abrupt thermal shock to the probe must be avoided at all times. After the probe is inserted completely, the initial status of the contact is checked by measuring the zero-bias conductance (ZBC) using the standard four-probe lock-in technique. If the resistance of the contact is too high, the piezoelectric bimorphs are driven to apply more pressure. Once the contact is confirmed to be stable, the temperature is slowly lowered by opening a needle valve between the liquid $^4$He reservoir and the sample space. At the same time, the driving voltage to the piezoelectric bimorphs is gradually increased to ensure a stable contact and any change of the ZBC is monitored by computer. When the temperature is close to T$_c$, the ZBC as a function of temperature is measured continuously with decreasing temperature. The temperature is controlled by adjusting the needle valve opening and the $^4$He pressure.

\subsection{$^3$He system}

The detailed procedure for making contacts to the sample when used with the $^3$He cryostat is as follows.
The sample and the gold tip are mounted as described in the previous section. The CAT rig is attached to the $^3$He cryostat which is fixed horizontally to a rigid frame. As described previously, the sample holder is thermally anchored to the $^3$He pot using multiple pairs of twisted Cu wires.
After the contact is formed using the mechanical screw, the frame holding the cryostat is carefully lifted to a vertical position and a vacuum can is attached to the 1 K stage using a silicone sealant. We stress that this is a delicate operation and the frame is required to provide rigidity to the CAT rig while lifting, and to protect the CAT rig from any mechanical contact. After the sealant is cured for one hour, the sample space, called the inner vacuum chamber (IVC) as shown in Fig.~\ref{fig:CATsetup},
is evacuated below 50 mTorr and back-filled with $^4$He exchange gas to 1 Torr. Finally, the $^3$He cryostat is slowly inserted into the VTI. During the initial cooling down, the sorption pump heater for the $^3$He pot is set to near 40 K. The bias for piezoelectric bimorphs is continuously adjusted during the cool-down to ensure that the contact remains stable. When the temperatures of both the 1 K stage and the $^3$He pot reach $\sim$10 K, they are kept at that temperature while the $^4$He exchange gas in the IVC is pumped to the low 10$^{-6}$ Torr range, which allows the 1 K stage to cool down further. When the temperature of the 1 K stage falls below 3 K, the $^3$He gas begins to condense. A complete condensation takes about 20 minutes. The sample temperature remains at around 1.2 K until the sorption pump heater power is lowered, thereby pumping on the $^3$He pot. A base temperature around 300 mK is typically attained. The sample temperature can be adjusted by controlling the sorption pump heater power.

\subsection{Cool-down Monitoring and Bimorph Adjustment}

From the shape of ZBC vs. temperature, one can identify the nature of the contact (tunneling-like, AR-like, or intermediate) by comparing the measured data with the conductance calculated using the BTK model. This information is useful to analyze and understand the conductance spectra. Once the base temperature is reached, the current vs. voltage and the conductance vs. voltage data are taken. These measurements are repeated for various magnetic fields and temperatures. The magnetic field is typically applied along the sample plane to minimize the demagnetization effect.

\section{\label{sec:performance}Performance and discussion}

We have carried out PCS measurements on three different superconductors, Nb and MgB$_2$ using the $^4$He cryostat, and CeCoIn$_5$ using the $^3$He cryostat.

\begin{figure}[b]
\includegraphics{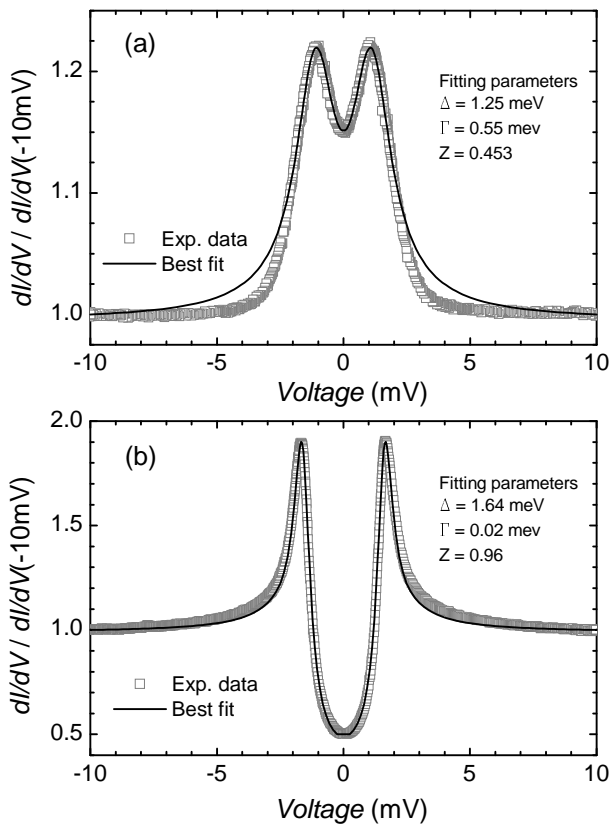}
\caption{\label{fig:NbPCS} Conductance vs. voltage characteristics for two Nb thin films grown in different systems, taken at (a) 1.41 K and (b) 1.32 K. The solid lines are the best fit curves using the BTK model with the fitting parameters as displayed. Note the conductance in (a) is best fit using more AR-like (smaller Z) parameters, while that in (b) is best fit using more tunneling-like (larger Z) parameters.}
\end{figure}

High quality Nb thin films are grown using dc magnetron sputtering in our group. Conductance spectra of two Nb thin films, deposited in different systems, are displayed in Fig.~\ref{fig:NbPCS}. The sample in Fig.~\ref{fig:NbPCS}(a) is 2100 \AA \ thick and exhibits the T$_c$ of 9.22 K and the residual resistivity ratio, defined as the ratio of resistivity at 300 K and at the onset temperature of the superconducting transition, of 115. The corresponding values for the sample in Fig.~\ref{fig:NbPCS}(b) are 4500 \AA, 9.27 K, and 66, respectively. Fig.~\ref{fig:NbPCS}(a) and (b) can be explained in terms of different surface
states of the two films. It is well known that various Nb oxides can form on the surface of as-grown
Nb thin films once exposed to the air or even kept inside a vacuum chamber.\cite{lindau74,grundner80} 
Therefore, the sensitivity of Nb to oxidation may well depend on the deposition system. For most of the thin films grown in one chamber, we have observed more tunneling-like conductance curves as shown in Fig.~\ref{fig:NbPCS}(b). The Nb thin film grown in the other chamber shows more AR-like curves, as plotted in Fig.~\ref{fig:NbPCS}(a). The conductance spectra can be analyzed by the BTK model using the three fitting parameters; the energy gap $\Delta$, the QP lifetime smearing factor $\Gamma$,\cite{dynes78}
and the effective barrier strength $Z$. The best fit curves are plotted as solid lines, and the fitting parameters, $\Delta=1.25\ meV$, $\Gamma=0.55\ meV$, and $Z=0.453$ for Fig.~\ref{fig:NbPCS}(a), and
$\Delta=1.64\ meV$, $\Gamma=0.02\ meV$, and $Z=0.96$ for Fig.~\ref{fig:NbPCS}(b). These results clearly show that the effective barrier strength is larger in (b). The Fermi velocities in Nb and Au are well matched,\cite{ashcroft76} which means the lower limit of $Z$ is nearly zero.\cite{blonder83} Thus, finite values of $Z$ can be attributed to the tunnel barrier layers on the surface. The energy gap for Fig.~\ref{fig:NbPCS}(a) is 0.25 meV smaller than 1.5 meV, the value for a bulk Nb. We also note that the fit curve deviates from the measured data above $\sim$2 mV. These results may be attributed to the proximity effect,\cite{pe} however, a detailed study is required to understand this behavior completely. Note also that the energy gap for Fig.~\ref{fig:NbPCS}(b)is 0.14 meV larger than the literature value. This observation is not unusual and we stress that Nb can have different superconducting properties depending on the cleanness of the material.\cite{niobium}

\begin{figure}[t]
\includegraphics{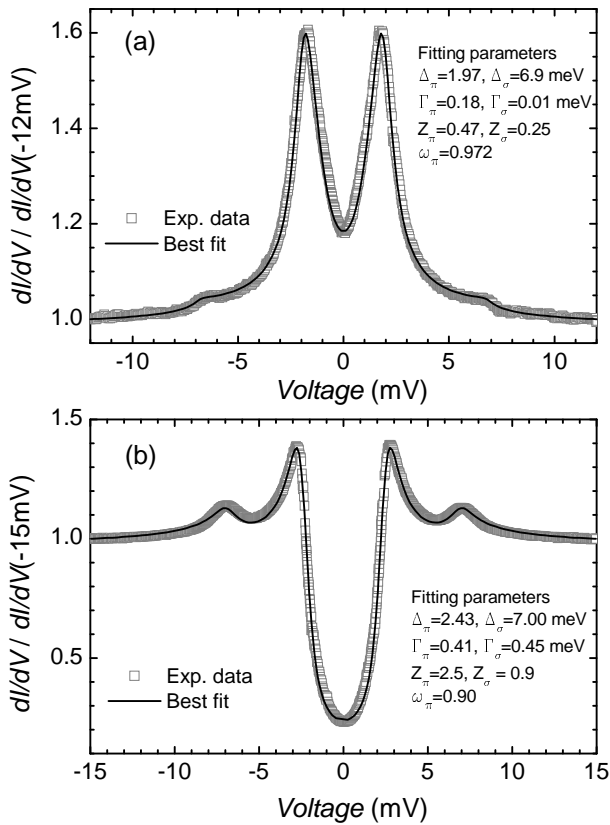}
\caption{\label{fig:MgB2PCS} Conductance vs. voltage characteristics for MgB$_2$ thin films. (a) PSU sample, taken at 1.46 K. (b) STI sample, taken at 1.38 K. The solid lines are best fit curves using a two-band BTK model with the fitting parameters as displayed. Note the conductance in (a) is best fit using more AR-like (smaller Z) parameters, while that in (b) is best fit using more tunneling-like (larger Z) parameters.}
\end{figure}

MgB$_2$ has drawn great deal of attention due to fundamental interest and application potential. It is recognized as the first superconductor which distinctly exhibits two gaps, as shown by a variety of different experiments\cite{szabo01,gonnelli02,bouquet01,iavarone02} with supporting theories.\cite{liu01,choi02} The point-contact spectra in Fig.~\ref{fig:MgB2PCS}(a) and (b) were taken on
MgB$_2$ thin films grown at the Pennsylvania State University (PSU) and at Superconductor Technology Inc. (STI), respectively. Both films are grown epitaxial with the $c$-axis normal to the substrates. T$_c$s are 40.5 K and 39.3 K for the PSU and STI films, respectively. Two peaks are seen in the conductance curves, reflecting two energy gaps in agreement with the literature. Similar to Nb thin films, the two MgB$_2$ samples show different characteristics: the conductance curve in Fig.~\ref{fig:MgB2PCS}(a) is more AR-like and that in Fig.~\ref{fig:MgB2PCS}(b)is more tunneling-like in nature. Again, this behavior can be attributed to different states on the film surface, depending on the film growth techniques.
MgB$_2$ is known to be an $s$-wave superconductor with strong electron-phonon coupling.\cite{liu01,choi02}
The Cooper pair condensates form on two disparate Fermi surfaces from three dimensional $\pi$- and two-dimensional $\sigma$-bands.\cite{liu01,choi02,kortus01} Unless the interband scattering is strong, the conductance data can be analyzed using a two-band BTK model with the total conductance given by the
sum of contributions from each band. In this case, there are seven fitting parameters with a weighting factor $w_\pi$ for the $\pi$-band as the seventh. The best fit curves are plotted as solid lines in Fig.~\ref{fig:MgB2PCS} with the fitting parameters displayed. For the data displayed in Fig.~\ref{fig:MgB2PCS}(a), $\Delta_\pi = 1.97 \textrm{meV}$, $\Delta_\sigma = 6.9 \textrm{meV}$, $\Gamma_\pi = 0.18 \textrm{meV}$, $\Gamma_\sigma = 0.01 \textrm{meV}$, $Z_\pi = 0.47$, $Z_\sigma = 0.25$,
and $w_\pi = 0.972$. As expected from the $c$-axis oriented texture of the film and the two-dimensional
nature of the $\sigma$-band Fermi surfaces, the signature for a large gap around 7 meV is much weaker, which is confirmed by the large value of $w_\pi$ required for the best fit. For Fig.~\ref{fig:MgB2PCS}(b), the best fit curve is obtained using $\Delta_\pi = 2.43 \textrm{meV}$,
$\Delta_\sigma = 7.00 \textrm{meV}$, $\Gamma_\pi = 0.41 \textrm{meV}$, $\Gamma_\sigma = 0.45 \textrm{meV}$, $Z_\pi = 2.5$, $Z_\sigma = 0.9$, and $w_\pi = 0.9$. In this case, two gaps are seen more clearly and confirmed by a larger weighting factor for the $\sigma$-band than for the data shown in Fig.~\ref{fig:MgB2PCS}(a). This observation can be explained by considering the texture of the film,
that is, imperfect alignment of grains along the $c$-axis. Values we obtain for the two energy gaps are in good agreement with the literature.\cite{szabo01,gonnelli02,iavarone02}

\begin{figure}[b]
\includegraphics{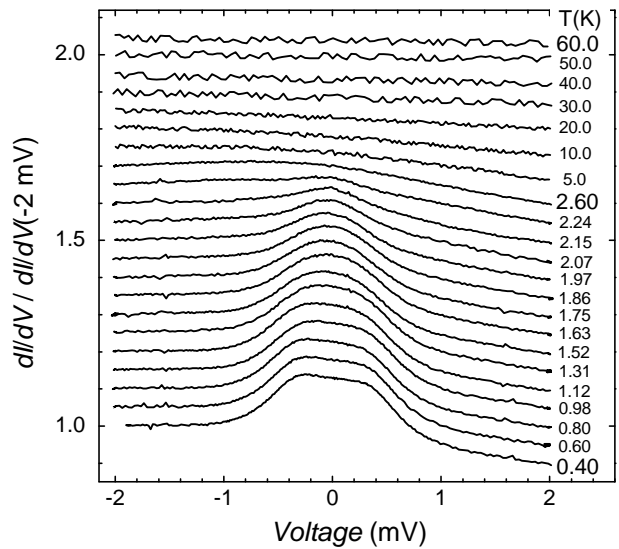}
\caption{\label{fig:Ce115PCS} Dynamic conductance spectra of a Au/CeCoIn$_5$ point contact between 60 K and 400 mK. Curves are shifted vertically by 0.05 for clarity. After Ref. 33.}
\end{figure}

\begin{figure}[t]
\includegraphics{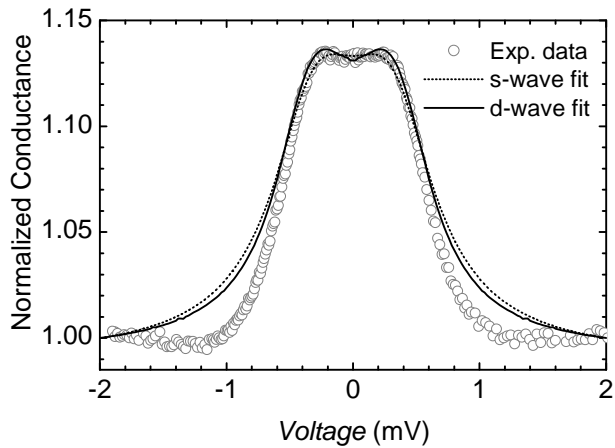}
\caption{\label{fig:Ce115PCS400mK} Conductance vs. voltage of a Au/CeCoIn$_5$ at 400 mK. The dashed(solid) line is a best fit curve using the extended BTK model based on $s$($d$)-wave pairing symmetry. After Ref. 33.}
\end{figure}

Our interest in the HFS, which have relatively low T$_c$s, requires a $^3$He cryostat or a dilution refrigerator to study their superconducting properties. In particular, we have studied CeCoIn$_5$ well below its T$_c$ (2.3 K) by attaching the CAT rig to the $^3$He cryostat. A CeCoIn$_5$ single crystal is etched in a concentrated hydrochloric acid to remove any residual indium. The point contact between a single crystal CeCoIn$_5$ and a gold tip is formed with the tip axis perpendicular to the crystal $ab$ plane so the current is injected along the $c$-axis. The dynamic conductance spectra, taken over a wide temperature range, from 60 K to 400 mK, are shown in Fig.~\ref{fig:Ce115PCS}.\cite{park05} An asymmetry in the background conductance is seen to develop starting at $\sim$ 40 K, which we attribute to the emergence of the coherent heavy-fermion liquid.\cite{nakatsuji04} This asymmetry is nearly constant below T$_c$ (2.3 K), where an enhancement of the sub-gap conductance is observed. A striking feature is the flat region in the subgap conductance at lower temperatures, reminiscent of AR. From estimations of the contact size and electronic mean free paths, the contact is shown to be in the Sharvin limit.\cite{park05} Analyses based on the extended BTK model\cite{kashiwayatanaka} indicate that $d$-wave symmetry is more likely than $s$-wave, in agreement with the literature.\cite{thompson03,izawa01,eskildsen03,aoki04} However, the full conductance curves cannot be fully accounted for, as shown in Fig.~\ref{fig:Ce115PCS400mK}. We claim that this failure is closely related to the heavily suppressed AR across a N/HFS interface,
\cite{gloos95,anders97} and provide a framework for better theoretical understanding.\cite{park05}

We have demonstrated that our CAT rig is a useful and reliable technique for the spectroscopic investigation of a variety of superconductors. Furthermore, the rig is stable and robust against thermal cycles. An obvious advantage of this technique is its versatility and its simplicity of implementation. 
We have shown it is adapted to both $^4$He and $^3$He cryostats. In addition, the tips can be exchanged so that different tip materials may be used, including ferromagnets and superconductors. An outstanding challenge remaining is the planar tunneling mode of the CAT rig. 

\section*{\label{sec:thank}ACKNOWLEDGEMENTS}

We are grateful to C. Gulyash and S. Schultz for their excellent machining work, J. L. Sarrao and J. D. Thompson at Los Alamos National Laboratory for providing CeCoIn$_5$ single crystals, B. Moeckly at Superconducting Technology Inc., J. M. Rowell at the Arizona State University, P. Orgiani and Q. Li at the Pennsylvania State University for providing MgB$_2$ thin films, J. Elenewski, B. F. Wilken, K. Parkinson, A. N. Thaler, P. J. Hentges, M. K. Brinkley, C. J. Ramsey, J. B. McMinis, A. O'Brien, X. Lu, and W. L. Feldmann for their experimental help. This work was supported by the U.S. Department of Energy, Division of Materials Sciences under Award No. DEFG02-91ER45439, through the Frederick Seitz Materials Research Laboratory and the Center for Microanalysis of Materials at the University of Illinois at Urbana-Champaign.

\end{document}